\documentclass[11pt,a4paper]{article}
\usepackage{amsmath,amssymb}
\usepackage{jheppubx}
\usepackage{graphicx}

\usepackage[usenames,dvipsnames]{xcolor}

\input epsf
\usepackage{epstopdf}
\usepackage{ upgreek }

\definecolor{darkgreen}{rgb}{0,0.4,0}
\definecolor{darkred}{rgb}{0.4,0,0}
\definecolor{darkblue}{rgb}{0,0,0.4}

\def\be{\begin{equation}}
\def\ee{\end{equation}}
\newcommand{\bea}{\begin{eqnarray}}
\newcommand{\eea}{\end{eqnarray}}

\input epsf

\newlength{\extraspace}
\newlength{\extraspaces}
\def\II{\relax{I\kern-.10em I}}

\def\IZ{\relax{\rm Z\kern-.34em Z}}
\def\IB{\relax{\rm I\kern-.18em B}}
\def\IC{{\relax\hbox{$\inbar\kern-.3em{\rm C}$}}}
\def\ID{\relax{\rm I\kern-.18em D}}
\def\IE{\relax{\rm I\kern-.18em E}}
\def\IF{\relax{\rm I\kern-.18em F}}
\def\IG{\relax\hbox{$\inbar\kern-.3em{\rm G}$}}
\def\IGa{\relax\hbox{${\rm I}\kern-.18em\Gamma$}}
\def\IH{\relax{\rm I\kern-.18em H}}
\def\II{\relax{\rm I\kern-.18em I}}
\def\IK{\relax{\rm I\kern-.18em K}}
\def\IP{\relax{\rm I\kern-.18em P}}

\def\inbar{\,\vrule height1.5ex width.4pt depth0pt}

\def\IR{\relax{\rm I\kern-.18em R}}

\def\lp10{\ell_p^{10}}
\def\lp11{\ell_p^{11}}
\def\R11{R_{11}}

\def\frac#1#2{{#1 \over #2}}

\newdimen\tableauside\tableauside=1.0ex
\newdimen\tableaurule\tableaurule=0.4pt
\newdimen\tableaustep
\def\phantomhrule#1{\hbox{\vbox to0pt{\hrule height\tableaurule width#1\vss}}}
\def\phantomvrule#1{\vbox{\hbox to0pt{\vrule width\tableaurule height#1\hss}}}
\def\sqr{\vbox{%
  \phantomhrule\tableaustep
  \hbox{\phantomvrule\tableaustep\kern\tableaustep\phantomvrule\tableaustep}%
  \hbox{\vbox{\phantomhrule\tableauside}\kern-\tableaurule}}}
\def\squares#1{\hbox{\count0=#1\noindent\loop\sqr
  \advance\count0 by-1 \ifnum\count0>0\repeat}}
\def\tableau#1{\vcenter{\offinterlineskip
  \tableaustep=\tableauside\advance\tableaustep by-\tableaurule
  \kern\normallineskip\hbox
    {\kern\normallineskip\vbox
      {\gettableau#1 0 }%
     \kern\normallineskip\kern\tableaurule}%
  \kern\normallineskip\kern\tableaurule}}
\def\gettableau#1 {\ifnum#1=0\let\next=\null\else
  \squares{#1}\let\next=\gettableau\fi\next}

 \def\eqnn#1{\xdef #1{(\secsym\the\meqno)}\writedef{#1\leftbracket#1}%
 \global\advance\meqno by1\wrlabeL#1}
 \def\eqna#1{\xdef #1##1{\hbox{$(\secsym\the\meqno##1)$}}
 \writedef{#1\numbersign1\leftbracket#1{\numbersign1}}%
 \global\advance\meqno by1\wrlabeL{#1$\{\}$}}
 \def\eqn#1#2{\xdef #1{(\secsym\the\meqno)}\writedef{#1\leftbracket#1}%
 \global\advance\meqno by1$$#2\eqno#1\eqlabeL#1$$}

\global\newcount\itemno \global\itemno=0

\def\itemaut#1{\global\advance\itemno by1\noindent\item{\the\itemno.}#1}

\def\({\left(}
\def\){\right)}

\usepackage[yyyymmdd,hhmmss]{datetime}
\newif{\ifeq}           
\eqtrue                 
\def\question#1
{
\ifeq
\textcolor{darkblue}{#1}
\fi
}

\nocite{*}

\begin{document}

\title{The Spinful Large Charge Sector of Non-Relativistic CFTs: From Phonons to Vortex Crystals}
\author{S.M. Kravec \&}
\author{Sridip Pal}  
\affiliation{Department of Physics, 
University of California at San Diego,
La Jolla, CA 92093} 

\emailAdd{skravec@ucsd.edu}
\emailAdd{srpal@ucsd.edu}

\abstract{
We study operators in Schr\"odinger invariant field theories (non-relativistic conformal field theories or NRCFTs) with large charge (particle number) and spin. Via the state-operator correspondence for NRCFTs, such operators correspond to states of a superfluid in a harmonic trap with phonons or vortices. Using the effective field theory of the Goldstone mode, we compute the dimensions of operators to leading order in the angular momentum $L$ and charge $Q$. We find a diverse set of scaling behaviors for NRCFTs in both $d=2$ and $d=3$.}
\maketitle

\section{Introduction and Summary}

Superfluid states of matter are one of most fundamental examples of spontaneous symmetry breaking and appear in countless systems from Helium-4 \cite{donnelly1991quantized,vitiello1996vortex,ortiz1995core,giorgini1996vortex} 
 to neutron stars \cite{baym1969superfluidity}. Superfluidity is also a possibility for finite density states of scale invariant critical systems \cite{hartnoll2016holographic}. Recently this observation has been used to perform explicit calculations of relativistic conformal field theory (CFT) data, despite strong coupling\cite{hellerman2015cft,hellerman2017note,monin2017semiclassics,banerjee2018conformal,de2018large,Jafferis:2017zna}. The key idea behind this is the fact that the large charge operators of the CFT correspond to finite density states on the sphere, which spontaneously break the conformal invariance and $U(1)$ corresponding to the charge. Superfluid phenomenology then becomes relevant for describing the large charge sectors of these CFTs. For example, another hallmark of superfluidity is the formation of vortices upon insertion of angular momentum. Therefore states with vortices correspond to large charge operators with spin, and calculating the energy of these vortices reveal the spining operator spectrum in CFT \cite{Cuomo:2017vzg}.
 
 
However, many interesting critical systems do not possess Lorentz symmetry. This includes ultracold fermi gases at ``unitarity", where observation of vortex lattices is perhaps the most dramatic evidence for a superfluid ground-state in a system which exhibits an emergent scale invariance\cite{zwierlein2005vortices}. At this critical point the system has a non-relativistic conformal symmetry, or Schr\"odinger symmetry. This symmetry algebra plays a pivotal role in understanding numerous physical systems\footnote{It is important to mention that Schr\"odinger symmetry is not simply the non-relativistic limit of the conformal symmetry but rather an entirely distinct algebra \cite{kobach:2018nmt, Henkel:2003pu}.}. Examples include the aforementioned ``fermions at unitarity"\cite{Regal:2004zza,Zwierlein:2004zz}, as well as systems comprised of deuterons \cite{Kaplan:1998tg,Kaplan:1998we}, ${}^{133}Cs$\cite{Chin:2001uan}, ${}^{85}Rb$ \cite{Roberts:1998zz},${}^{39}K$ \cite{loftus2002resonant}, and various spin chain models \cite{Chen:2017tij}. There has been significant progress in understanding the consequences of Schr\"odinger symmetry and its realization in field theory.\cite{Mehen:1999nd,Nishida:2010tm,Nishida:2007pj,goldberger2015ope,Golkar:2014mwa,Pal:2018idc} These non-relativistic conformal field theories (NRCFTs) admit a state-operator correspondence akin to their relativistic cousins. Operators with ``particle number" charge are related to states in a harmonic potential.\cite{nishida2007nonrelativistic} This has been exploited to calculate the energies of few-body quantum mechanics systems in a harmonic trap. This correspondence also implies a way that the spectrum of NRCFTs can be determined. The operators with large charge correspond to finite density states in the trap. These states of matter sometimes admit a simple effective field theory description, enabling semi-classical calculations controlled in the large charge limit \cite{Kravec:2018qnu,Favrod:2018xov}.

The simplest and most physically relevant possibility is that of a superfluid ground-state, which is the situation we will explore here.\footnote{It should be emphasized that this is not the only possibility. Ultimately the question ``Given this NRCFT, what state of matter describes its large charge sector?" depends on the NRCFT, which we treat as UV physics. However we expect our results to be valid for a wide set of NRCFTs, including some of physical relevance such as unitary fermions \cite{zwierlein2005vortices}.} Extending upon the results of \cite{Cuomo:2017vzg,Kravec:2018qnu}, we study NRCFT operators which have both large charge and spin. Such operators correspond to either phonon or vortex excitations of the superfluid. We then compute the leading order scaling of their dimensions $\Delta_{Q,L}$ as functions of their angular momentum $L$ and number charge $Q$ and find a diverse range of behaviors in $d=2$ and $d=3$.

\subsection*{Trailer of the Results:}
We compute the leading scaling dimension $\Delta_{Q,L}$ of spinining operators of a non-relativistic conformal field theory as a function of $U(1)$ charge $Q$ and angular momentum $L$ in the large charge limit. The answers are determined up to a single Wilson coefficient $c_0$ in the EFT description. We leverage the state operator correspondence to arrive at the result that depending on the range of angular momentum, the spinning operators correspond to different excitation modes of the superfluid. For a smaller range of angular momentum, we find that they correspond to phonon with angular momentum $L$. As we increase the angular momentum, we pass through a regime where a single vortex becomes energetically favorable. If we further increase the angular momentum, multiple vortices develop and the superfluid exhibits an effective ``rigid body motion" where we can neglect the discrete nature of the vortices. 

In $d=2$, the leading behavior has $3$ regimes and is given as follows:
\be \label{master-d2}
 d=2 ~~~~ \Delta_{Q,L} = \begin{cases} \sqrt{L} + \Delta_Q ~~~~~~& 0<L\leq Q^{1/3} \\\\
  \sqrt{\frac{c_0\pi}{2}} \sqrt{L} \log L + \Delta_Q ~~~~~ &Q^{1/3} < L \leq Q\\\\
\sqrt{\frac{9c_0\pi}{2}} \left( \frac{L^2}{Q^{3/2}}\right)+\Delta_Q ~~~~~ &Q < L <Q^{3/2}
  \end{cases}
\ee
where $\Delta_Q=\frac{2}{3} \left(\frac{1}{\sqrt{2\pi c_0}}\right) Q^{3/2}$ is the contribution from ground state energy in $d=2$.\\ 

In $d=3$ dimensions, we have $4$ regimes, given by:
\be \label{master-d3}
 d=3 ~~~~ \Delta_{Q,L} =\begin{cases} \sqrt{L} + \Delta_Q~~~~~~ &0<L\leq Q^{2/9} \\\\
\alpha \left(\frac{L}{Q^{1/9}}\right) + \Delta_Q~~~~~~ &Q^{2/9}<L\leq Q^{1/3}  \\\\
 \left(\frac{5\pi^4 c_0}{8\sqrt{2}}\right)^{1/3}  L^{2/3}\log L+ \Delta_Q~~~~~~ &Q^{1/3} < L \leq Q\\\\
\frac{1024}{25}\left(\frac{32c_0^2}{25\pi^4}\right)^{1/6}\left(\frac{L^2}{Q^{4/3}}\right) +\Delta_Q ~~~~~ &Q< L < Q^{4/3}
 \end{cases}
\ee
where $\Delta_{Q}=\frac{3}{2}\frac{1}{\sqrt{2\pi}}\left(\frac{6}{15\sqrt{\pi}c_0}\right)^{1/3} Q^{4/3}$  is the contribution from ground state energy in $d=3$ and $\alpha$ is an undetermined $\mathcal{O}(1)$ coefficient. We make two remarks at this point. The first one is that while for $d=2$, the transition happens from a single phonon regime to vortex regime at $L\sim Q^{1/3}$, for $d=3$, there is a regime $Q^{2/9}\leq L \leq Q^{3/9}$, where neither vortex nor the single phonon solution gives the lowest energy. It is a cross-over describing the physics of a vortex string forming near the boundary of the trap where our EFT is strongly coupled. The only well defined configuration in this angular momentum regime contains multiple phonons, and we determine the scaling from that. The second remark is that the EFT description breaks down whenever $\Delta_{Q,L}-\Delta_Q \sim \Delta_{Q}$ , so we can not probe operators with larger angular momentum with this method. 

The rest of the paper is organized as follows. We briefly review the superfluid hydrodynamics and large charge NRCFT in section~\ref{sec:review}. The section~\ref{sec:phonons} details out the contribution coming from phonons and derives the regime where it is energetically favorable to have them. Subsequently, we discuss the single vortex in $d=2$ and $d=3$ in section~\ref{sec:vortex}. The multi-vortex and rigid body motion is elucidated in section~\ref{sec:multi-vortex}  followed by a brief conclusion and future avenues to explore in section~\ref{sec:conclusion}. Some of our results and validity regimes are more apparent in dual frame using particle-vortex duality which we elaborate on in appendix~\ref{app:A}. The appendix~\ref{app:B} contains a contour integral useful for calculating interaction energy of multiple vortices in $d=3$.


\section{The set up: Superfluid Hydrodynamics and Large Charge NRCFT}\label{sec:review}

In this section we briefly review the superfluid hydrodynamics in the Hamiltonian formalism, specialized to the case of a Schr\"odinger invariant system in a harmonic potential $A_0=\frac{1}{2}\omega^2 r^2$. All of our results will be to leading order in the derivative expansion. For a more in-depth review of the formalism, we refer to \cite{son2006general,Kravec:2018qnu,Favrod:2018xov}. 

The low-energy physics of a superfluid is determined by a single Goldstone field $\chi$. The leading order Lagrangian determines the pressure of the system:
\be \label{leading-order-lagrange}
\mathcal{L}= c_0 X^{\frac{d+2}{2}} \equiv P(X) ~~~~~ X\equiv \partial_0 \chi -  A_0 - \frac{1}{2}(\partial_i \chi)^2
\ee
The number density and superfluid velocity are defined respectively as:
\be \label{number-velocity-def}
n= \frac{\partial L}{\partial \dot{\chi}}= c_0 \left(\frac{d}{2}+1\right) X^{\frac{d}{2}} ~~~~~ v_i = - \partial_i \chi
\ee
The action \eqref{leading-order-lagrange} has a $U(1)$ symmetry of $\chi \rightarrow \chi + c$ whose current can be written as:
\be \label{u1-current}
j^\mu = \( n , n v^i \)
\ee

The Hamiltonian density comes out to be:
\be \label{leading-order-hamiltonian}
\mathcal{H} = n \dot{\chi} - \mathcal{L} = n \left(X + A_0 + \frac{1}{2}v^2 \right) - P(X) 
\ee
Now, using the thermodynamic relation $nX - P(X) \equiv \epsilon(n)$:
we can simplify \eqref{leading-order-hamiltonian} and express the Hamiltonian as:
\be \label{hamil2}
H=\int \text{d}^dx ~\mathcal{H} ~~~~~\mathcal{H} = \frac{1}{2}n v^2 + \epsilon(n) + n A_0
\ee
.

Note that the presence of the harmonic trap implies the density is non-uniform and vanishes at radius $R_{TF} = \sqrt{\frac{2\mu}{\omega^2}}$. For most values of $r$ the density is large and varies slowly compared to the UV length scale $\frac{1}{\sqrt{\mu}}$. However, the large charge expansion begins to break down at $R^*=R_{TF}-\delta$ where $\delta \sim \frac{1}{(\omega^2 \mu)^\frac{1}{6}}$\cite{Kravec:2018qnu,son2006general}. There is a boundary layer of thickness $\delta$ where the superfluid effective field theory (EFT) cannot be trusted as it is no longer weakly coupled. At leading order in the derivative expansion this does not effect the observables but leads to divergences at higher orders.\footnote{These are UV divergences which can be canceled by counter-terms localized at this edge, as suggested by Simeon Hellerman in a private communication.}

Given this set up, the ground-state at finite density corresponds to the classical solution of $\chi_{cl}=\mu t$. The number charge of this configuration is determined from $\mu$ by:
\be \label{number-charge-cl}
Q \equiv \int \text{d}^dx ~n_{cl}(x) = c_0 \left(\frac{d}{2}+1\right) \int \text{d}^dx ~\left(\mu - A_0\right)^\frac{d}{2} = \frac{1}{\xi} \left(\frac{\mu}{\omega}\right)^d
\ee
where $\frac{1}{c_0}=\frac{\Gamma(\tfrac{d}{2}+2)}{\Gamma(d+1)}(2\pi \xi^2)^{\tfrac{d}{2}}$. We can then compute the ground-state energy as function of $Q$ using \eqref{hamil2}:
\be \label{ground-state-energy}
E_Q = \int \text{d}^dx ~[\epsilon(n_{cl}) + n_{cl} A_0] = \omega\xi\left(\frac{ d}{d+1}\right) Q^{\tfrac{d+1}{d}}
\ee

Via the state-operator correspondence of NRCFTs, this semi-classical calculation determines the dimension of a charged scalar operator to leading order in $Q$ as $\Delta_Q = \frac{E_Q}{\omega}$.  In particular, we have obtained \cite{Kravec:2018qnu}:
\begin{align}
\Delta_Q=\begin{cases}
\frac{2}{3} \xi Q^{3/2} \quad \text{for}\ ~~~ d=2\\
\frac{3}{4}\xi Q^{4/3} \quad \text{for}\ ~~~ d=3
\end{cases}
\end{align} 

In this work, we'll be interested in excited state configurations which carry some angular momentum. These will correspond to spinful operators in the large charge sector of the NRCFTs which the superfluid EFT describes. The simplest of these excitations are phonons; smooth solutions of the equation of motion with $\chi_{cl} = \mu t + \pi$. Expanding $\pi$ in modes $\pi_{n,\ell}$, the Hamiltonian can be written to leading order in the derivative expansion as:
\be \label{hamil-phonons}
H = H_0 + \sum_{n,\ell} \omega(n,\ell) \pi^\dag_{n,\ell} \pi_{n,\ell} + \cdots
\ee
where $\omega(n,\ell)$ is the dispersion relation for phonons:
\be \label{phonon-dispersion}
\omega(n,\ell) = \omega \left(\frac{4}{d}n^2 + \left(4-\frac{4}{d}\right)n + \frac{4}{d}n \ell + \ell\right)^{\frac{1}{2}}
\ee
for $n$ is a positive integer and $\ell$ is the total angular momentum. The phonon wavefunctions are given as $f_{n,\ell} \sim (\frac{r}{R_{TF}})^\frac{\ell}{2} G_{n,\ell}(r) Y_{\ell}$ where $G_{n,\ell}$ is a hypergeometric function and $Y_\ell$ is a spherical harmonic. A state with $M$ phonon modes of $\{n=0,\ell=1\}$ can be identified as the descendant operator $\vec{\partial}^M \mathcal{O}_Q$ with dimension $\Delta_Q +M$. Additionally, NRCFTs have another generator of descendants $\partial_t$ which corresponds to the phonon with $\{n=1,\ell=0\}$. States that can be created by adding phonons with other values of $n$ and $\ell$ correspond to distinct  primaries \cite{Kravec:2018qnu}.\footnote{They are primary as they are by construction annihilated by the lower operators $K$ and $C$ which correspond to $\pi_{n=0,\ell=1}$ and $\pi_{n=1,\ell=0}$ respectively.}

The other configuration of a superfluid that can support angular momentum is a vortex, which gives rise to a singular velocity field of the condensate. This is a distinct semi-classical saddle point which is not simply related to the ground state. It must therefore correspond to a unique set of spinful charged operators present in all NRCFTs whose scalar large charge sector is described by the superfluid EFT.

These two excitations, phonons and vortices, are the configurations of the superfluid we know support angular momentum. In the rest of the paper we answer the question, what is the lowest energy configuration of the superfluid for a given angular momentum? By answering this and using the superfluid EFT defined above we compute the scaling behavior of operators carrying charge and angular momentum.

\section{Phonons}\label{sec:phonons}

The simplest excited state(s) with angular momentum are phonons. From the dispersion \eqref{phonon-dispersion}, we can see that the lowest energy configuration with angular momentum $L$ is a single phonon with $n=0$ and $\ell=L$. This is known as a ``surface mode" as the wavefunction is nodeless and supported mostly at the end of the trap. The energy cost of this single phonon is given by\footnote{Note this is parametrically lower in energy than in the relativistic case studied in \cite{Cuomo:2017vzg}, as the phonon spectrum on the sphere is $\epsilon(\ell) =\sqrt{ \frac{1}{2}\ell(\ell+1)}$.}:
\be \label{deltaE-phonon}
\Delta E = \omega L^{\frac{1}{2}}\,.
\ee
However the validity of \eqref{phonon-dispersion} rests on the assumption that the phonon modes do not carry large amounts of momentum. In particular, the surface mode wavefunction has $f_{ \ell} \sim (\frac{r}{R})^\frac{\ell}{2} Y_{\ell}$ which for large $\ell$ is increasingly concentrated at the edge of the trap. Once the support of the phonon wavefunction is mostly within the boundary region of thickness $\delta$, we can no longer trust the solution or the dispersion \eqref{phonon-dispersion}. This occurs when $\frac{R_{TF}}{\ell}$ becomes comparable to $\delta$ \cite{pethick2008bose}. This yields a maximum angular momentum for phonons: $\ell_{max} \sim Q^{\frac{2}{3d}}$. 

Thus we have the following scalings for operator dimensions:
\be \label{single-phonon-op-scaling-d2}
 d=2 ~~~~ \Delta_{Q,L} = L^{\frac{1}{2}} + \Delta_Q ~~~~~~ 0<L\leq Q^{\frac{1}{3}} 
\ee
\be \label{single-phonon-op-scaling-d3}
 d=3 ~~~~ \Delta_{Q,L} = L^{\frac{1}{2}} + \Delta_Q ~~~~~~ 0<L\leq Q^{\frac{2}{9}} 
\ee
where $\Delta_Q$ is the operator dimension determined from \eqref{ground-state-energy}. 

We can also consider multi-phonon configurations and ask ourselves whether it is energetically favorable to have a single phonon rather than multi phonon configuration, given total angular momentum. In order to answer this, we assume that phonon interactions are negligible, suppressed to leading order in the $Q$-expansion, so the energy and angular momentum of multiple phonons add linearly. In particular, suppose we have $N_\gamma$ phonons, each carrying angular momentum $\ell$. The energy and angular momentum to leading order is:
\be \label{energy-angular-multi-phonon}
\Delta E = \omega N_\gamma\ell^{\frac{1}{2}} ~~~~~ L = N_\gamma \ell
\ee

This tells us that for a given angular momentum $L$, it is energetically favourable to have a single phonon carrying the entire angular momentum rather than multiple phonons carrying it altogether.\footnote{One can also arrive at the same conclusion by considering $N_\gamma$ phonons, each carrying angular momentum $\vec{\ell_i}$. The energy and angular momentum to leading order is then given by:
\be \label{energy-angular-multi-phonon-modified}
\Delta E = \omega \sum_i|\vec{\ell}_i|^{\frac{1}{2}} ~~~~~ L = \left|\sum_i \vec{\ell}_i\right|
\ee
We have 
\begin{align}
\Delta E = \omega \sum_i|\vec{\ell}_i|^{\frac{1}{2}}= \omega\sqrt{\sum |\vec{\ell}_i|+\sum\sqrt{|\vec{\ell}_i||\vec{\ell}_j|}} \geq \omega\sqrt{\left|\sum \vec{\ell}_i\right|+\sum\sqrt{|\vec{\ell}_i||\vec{\ell}_j|}} \geq \omega \sqrt{L}
\end{align}
Hence, the minimum value is obtained when all the $l_i=0$ except one i.e. we land up with single phonon case. On the other hand, if all the $\vec{\ell}_i$'s are along same direction, then using Cauchy-Schwartz inequality, one can obtain $\Delta E \leq \omega (N_\gamma)^{1/2}\sqrt{L}$, which implies that the energy would be maximized if each phonon carries angular momentum of $L/N_\gamma$.
}

As we'll see below, naively a single phonon of $\ell = L$ would always be the most energetically favorable configuration per angular momentum. However the cutoff of  $\ell_{max} \sim Q^{\frac{2}{3d}}$ means we cannot trust this conclusion beyond $L=\ell_{max}$. Multi-phonon configurations are in principle valid for larger values of $L$.\footnote{$N_{\gamma}$ cannot be made arbitrarily large as the assumption that phonon interactions are suppressed breaks down.} The most energetically favorable of which has $N_{\gamma}$ phonons with $\ell=\ell_{max}$, which gives the scaling:
\be \label{multi-phonon-best-scaling}
\Delta E = \omega L \ell_{max}^{-1/2} ~~~~~ \ell_{max} \sim Q^{\frac{2}{3d}}
\ee
where we cannot determine the dimensionless coefficient from $\ell_{max}$ as it depends on how we regulate the cutoff region of size $\delta$. Nevertheless, the linear scaling in $L$ means we can compare to other configurations such as vortices. In particular, we will arrive at the conclusion that whenever $L\geq Q^{1/3}$, the minimum energy configuration with a given angular momentum starts to be attained by vortex solutions.


For $d=2$, the transition happens from a single phonon regime to vortex regime at $L\sim Q^{1/3}$, while for $d=3$, there is a regime $Q^{2/9}\leq L \leq Q^{1/3}$, which is inaccessible by both the vortex string and the single-phonon configurations. The most energetically favorable configuration, consistent within the leading order EFT analysis, is therefore the multi-phonon configuration above with a macroscopic number of phonons $N_\gamma \sim Q^{\frac{1}{9}}$ at the upper bound $L \sim Q^{\frac{1}{3}}$. 

This would imply the following scaling for the operator dimension:
\be \label{multi-phonon-op-scaling-d3}
 d=3 ~~~~ \Delta_{Q,L} = \alpha L Q^{-1/9} + \Delta_Q ~~~~~~ Q^{2/9}<L\leq Q^{1/3} 
\ee
where $\alpha$ is an unknown order one coefficient.

However the exact nature of this state appears to be related to UV physics of how a vortex string configuration forms from surface mode phonons in the boundary region of the condensate, which is inaccessible within our formalism. We therefore cannot give a full accounting of this regime of angular momentum. Beyond $L > Q^{\frac{1}{3}}$ we can be confident the lowest energy configuration is a vortex, as we'll now discuss.

\section{Single Vortex in the Trap}\label{sec:vortex}

A vortex is a configuration of the superfluid with a singular velocity field carrying angular momentum. The singular nature arises because of the relation \eqref{number-velocity-def} implying that $v_i$ is necessarily irrotational except due to defects in the field $\chi$; configurations where $\int_{\mathcal{C}} \text{d}\chi = 2\pi s$ for some integer $s$. In $d=2$ these are particle like excitations while in $d=3$ they correspond to strings, these will be the dimensions we focus on in this work. In fact this language can be made precise via particle-vortex duality, where vortices are ``charged" objects under some dual gauge field. Adapting this duality to the Schr\"odinger invariant superfluid has been done in Appendix \ref{app:A} but it is inessential for describing the leading order results.

The simplest configuration in the trap is a single static vortex for which the condensate order parameter changes by only $2\pi$.\footnote{This is in contrast to the CFT case where a minimum of two vortices are needed on the sphere to ensure compatibility with the Gauss law.} The approximate velocity profile $v_i$ of such a configuration is:
\be \label{velocity-approximate}
v_i = \frac{\epsilon^{ij}(r_j-R_j)}{(\vec{r}-\vec{R})^2} 
\ee
where $r$ is the radial coordinate in $d=2$ or the axial coordinate in $d=3$, and $\vec{R}$ is the location of the vortex and we assume that the vortex is streched along the $z$ axis.

The presence of the vortex changes the semi-classical number density, making it singular at $r=R$. Before that point the density vanishes, implying a short distance cutoff for the superfluid EFT. This is the `vortex core size' $a$ whose scaling dimension we can determine as follows. 

One interpretation of the non-uniform density \eqref{number-charge-cl} is that the effective chemical potential is distance dependent. In the presence of a vortex at $\vec{r}=\vec{R}$ it is given as:
\be \label{chem-eff-position}
\mu_{eff}(\vec{r}) \equiv \mu - \frac{1}{2}\omega^2 r^2 - \frac{1}{2}\frac{1}{|\vec{r}-\vec{R}|^2}
\ee
This determines a locally varying UV length scale $\frac{1}{\sqrt{\mu_{eff}}}$. The EFT, which is controlled in the limit of large density, becomes strongly coupled at the length $a$ when $a \sim \frac{1}{\sqrt{\mu_{eff}}}$. Solving this equation for $a$ gives the scaling relations\footnote{This is an equivalent condition to cutting off the theory when the velocity field sourced by the vortex becomes comparable to the local speed of sound in the superfluid $c_s^2 \sim \frac{\partial P}{\partial n} \sim X$.}
\be \label{cutoff-improved}
d=2 ~~~ a \sim \frac{1}{\sqrt{\mu}} \frac{1}{\sqrt{1-\frac{R^2}{R_{TF}^2}}} ~~~~~~ d=3 ~~~ a \sim \frac{1}{\sqrt{\mu}} \frac{1}{\sqrt{1-\frac{R^2}{R_{TF}^2}- \frac{z^2}{R_{TF}^2}}}
\ee 
Near the center of the trap, $a$ is on order the UV length scale $\frac{1}{\sqrt{\mu}}$. However as the vortex approaches the boundary of the trap, either in its placement $\vec{R}$ or along the length of the vortex string in $d=3$, the fact the density is depleted due to the trap implies the cutoff near the vortex string must happen sooner \cite{bruun2001vortex}. As mentioned previously, the EFT is already strongly coupled in the boundary region of size $\delta$. Therefore the largest placements of the vortex we can confidently study have $R = R_{TF}-\delta$ where the core size scales as $a \sim \frac{1}{(\mu \omega^2)^{\frac{1}{3}}}$ which is still parametrically suppressed in $\mu$.

Regulating this divergence as described above, the correction to the semi-classical number density due to the vortex is subleading in $\mu$ and therefore negligible for leading order results. This implies the dominant contribution to the energy of a vortex configuration comes from the kinetic energy of the velocity field. 


The velocity field \eqref{velocity-approximate} does not define a stationary flow in the sense that $\partial_i(n v^i) \neq 0$ because of the inhomogeneity of the density. This inhomogenity will cause the vortex to precess in a circle \cite{sheehy2004vortices}. However since the density varies slowly, as previously discussed, the correction to the velocity field due to this is suppressed in the large-charge expansion.  Using particle-vortex duality, this is equivalent to the assumption that particle sourcing the gauge field in dual description has suppressed velocity, hence we are effectively dealing with an electrostatic scenario. The details are relegated to the appendix \ref{app:A}, in particular, the discussion after \eqref{3d-gauss}. 

We remark that in dual frame, the cloud boundary is like a conductor, hence the tangential electric field should be vanishing. This means the velocity field of the vortex should be such that there is no radial outflow of particles out of the trap. Given this condition, one might worry that the velocity field above does not vanish at the boundary $R_{TF}$. However, since we require the normal component of the flow to vanish at the boundary i.e. $\hat{N}\cdot (n \vec{v})=0$ where $\vec{N}$ is a vector normal to the trap at boundary, the inhomogeneity of the superfluid comes to rescue and the condition is trivially met by the vanishing of the density $n(x)$ at $R_{TF}$ \cite{groszek2018motion}. \footnote{This is generically known as a ``soft boundary". Had we been dealing with homogenous fluid with non vanishing density at boundary, we ought to consider a mirror vortex configuration to ensure the imposition of $\hat{N}\cdot (n \vec{v})=0$, this is just like considering the mirror charge while solving for electric in the presence of a conductor. Regardless, such modifications to the velocity field in the boundary region of the inhomogenous condensate give suppressed corrections to our leading order results below.} 

In what follows, we will be evaluating the energy and angular momentum of vortex configurations in $d=2$ and $d=3$ spatial dimensions.

\subsection{Single vortex in $d=2$}
Let's first work in $d=2$ with the velocity field given by \eqref{velocity-approximate}. The difference in energy between the vortex state and the ground state can then be computed from the kinetic energy of Hamiltonian \eqref{hamil2} as:
\begin{align} \label{energy-integral-2d-vortex}
\Delta E &= \int \text{d}^2x ~ \frac{1}{2}n v^2 = c_0 \mu \int \text{d}^2x ~\left(1-\frac{r^2}{R_{TF}^2}\right) \frac{1}{(\vec{r}-\vec{R})^2}
\end{align}
As mentioned, there is a divergence at $r=R$ which we will regulate by assuming a vortex core size of $a(R) \sim \frac{1}{\sqrt{\mu_{eff}}}$ where $\mu_{eff}=\mu\left(1-\tfrac{R^2}{R^2_{TF}}\right)$. Evaluating the integral \eqref{energy-integral-2d-vortex}  gives:
\begin{align} \label{energy-result-2d-vortex}
\Delta E &= 2c_0 \pi \mu \left(1-\frac{R^2}{R_{TF}^2}\right)\left[\log\left(\frac{R_{TF}}{2a(R)}\right)+\frac{1}{2}\log\left(1-\frac{R^2}{R_{TF}^2}\right)-1\right]+c_0\pi \mu+O(a)
\end{align}

We can also compute the angular momentum  via the integral:
\begin{align} 
\vec{L} &= \int \text{d}^2x~n \vec{v}\times \vec{r} 
\end{align}
For our configuration the angular momentum is entirely in the $\hat{z}$ direction with magnitude:
\begin{align} \label{angular-momentum-integral-vortex-2d}
 L= 4\pi c_0 \mu \int_{R}^{R_{TF}} \text{d}r~ r \left(1-\frac{r^2}{R_{TF}^2}\right) = 2\pi c_0 \frac{\mu^2}{\omega^2} \left(1- \frac{R^2}{R_{TF}^2}\right)^2 
 \end{align}
where we've used $\oint_r \vec{v}\cdot d\vec{\ell}  = 2\pi$ for a circle centered at the origin of radius $r>R$, and otherwise vanishes.

As one can see, it is energetically favorable for the vortex to appear at the edge of the cloud $R \approx R_{TF}$. However we cannot trust the solution in the regime of low density near there for reasons previously discussed. Therefore the largest distance the vortex can be where we have confidence in the validity of the semi-classical approximation is $R^*=R_{TF} - \delta$. This gives a minimum angular momentum, of the vortex configuration $L_{min} \sim Q^{\frac{1}{3}}$. The largest value of the angular momentum occurs when the vortex is in the center at $R=0$ with $L_{max} \sim Q$.

Combining these results gives the leading order expressions for the operator dimensions in terms of $L$ and $Q$ as:



\be \label{2d-op-dim-single-vortex}
d=2 ~~~~ \Delta_{Q,L}= \sqrt{\frac{c_0\pi}{2}} \sqrt{L} \log L + \Delta_Q ~~~~~ Q^{\tfrac{1}{3}} < L \leq Q
\ee


\subsection{Single vortex in $d=3$}
Let's consider the case of $d=3$ now. The minimal energy excitation is a single vortex string. The string must necessarily break the spherical symmetry of the trap. We will consider the string being streched along the $z$-axis, ensuring that all the angular momentum is $L = L_z$.\footnote{A curved string will generically have to be longer in order to carry the same angular momentum, as parts of the velocity field it sources will cancel against each other. The longer strings will be energetically more expensive, making the straight line configuration energetically favorable to leading order.}

The energy of the vortex string again comes from the kinetic energy and can be evaluated as:
\be \label{energy-vortex-string}
\Delta E = \int \text{d}^3x ~ \frac{1}{2}n v^2 = \int_{-Z(R)}^{Z(R)}\text{d}z ~ T(z,R)
\ee
where $T(z,R)$ is the tension of the string and $Z(R) = R_{TF} \sqrt{\left(1- \frac{R^2}{R_{TF}^2}\right)}$ defines the integration bound along the length of the string.

The tension can be computed via a similar integral in $d=2$ as:
\begin{align} 
\nonumber T(z,R) &= \frac{1}{2} \int_0^{r(z)} \text{d}r r \int_0^{2\pi} \text{d}\phi ~ n(r,z) \frac{1}{(\vec{r}-\vec{R})^2} \\
\label{tension-integral}
&= \pi n(R,z)\left[ \log \left(\frac{r(z,R)}{a(z,R)}\right)-\log \left(1+\frac{r(z)}{r(z,R)}\right)\right]+\cdots
\end{align}
where $\cdots$ refer to the non logarthimic pieces. Here $n(r,z)=\frac{5}{2}c_0 \mu^{\frac{3}{2}}\left(1-\frac{1}{R_{TF}^2}(r^2+z^2)\right)^{\frac{3}{2}}$ is the number density, $r(z) = R_{TF}\sqrt{1-\frac{z^2}{R_{TF}^2}}$ is the radial (radius in cylindrical co-ordinate) size of the trap at a height $z$ and  $r(z,R) = R_{TF}\sqrt{1-\frac{z^2+R^2}{R_{TF}^2}}$.
Integrating the leading logarthimic piece along the string length gives the energy:
\begin{align} 
\nonumber \Delta E&= \int_{-Z(R)}^{Z(R)} \text{d}z\ \pi n(R,z)\log \left(\frac{r(z,R)}{a(z,R)}\right)\\
\label{vortex-string-energy}
&=\frac{15}{16} \pi ^2 c_0 \mu ^{3/2} R_{TF} \left(1-\frac{R^2}{R_{TF}^2}\right)^2 \left[\log \left(1-\frac{R^2}{R_{TF}^2}\right)+ \log \left(R_{TF}\sqrt{\mu}\right) \right]
\end{align}

Evaluating the angular momentum of this configuration is similar to $d=2$ and yields:
\begin{align} \label{angular-momentum-vortex-3d}
L &=\int_{-Z(R)}^{Z(R)} \text{d}z\ \int_{R}^{r(z)} \text{d}r\ r \left[\frac{5}{2}c_0\mu^{\frac{3}{2}} \left(1-\frac{r^2+z^2}{R^2_{TF}}\right)^{\frac{3}{2}}\right]=\frac{5 \pi c_0}{8\sqrt{2}}  \left(\frac{\mu}{\omega}\right)^{3}  \left(1-\frac{R^2}{R_{TF}^2}\right)^3
\end{align}

Again the lowest allowed value of the angular momentum occurs for a vortex at $R^*=R_{TF}-\delta$ and scales as $L_{min}\sim Q^{\frac{1}{3}}$ while the maximum occurs at $R=0$ with $L_{max} \sim Q$.

Together these results imply the scaling:
\be \label{3d-op-dim-single-vortex}
d=3 ~~~~ \Delta_{Q,L} =\left(\frac{5\pi^4 c_0}{8\sqrt{2}}\right)^{1/3}  L^{2/3}\log L+ \Delta_Q ~~~~~~ Q^{\frac{1}{3}} < L \leq Q
\ee

This determines the leading order dimension for the operator which creates the vortex string but we can also study the spectrum of operators above it. For example, the presence of a vortex string along the $\hat{z}$-direction should split the phonon $m$ degeneracy in \eqref{phonon-dispersion}. Treating this perturbatively, such a splitting is suppressed in the charge\footnote{One could also consider the energy of a vortex-phonon configuration. The ``interaction energy" between the two is given as $\int \text{d}^dx ~ \vec{v}_{vortex} \cdot \vec{\partial} \pi$ which is also suppressed in the $Q$ expansion.} $Q$ \cite{bruun2001vortex}.

Besides phonons, there are unique excitations of the vortex string related to displacements of position. These are known as ``Kelvin modes" and they define another set of low-lying operators above the one which created the vortex string. These modes are basically the radial displacement of the vortex core from the original axis.
For long wavelength modes $k a(z,R) \ll 1$ and in the regime where $z \ll R_{TF}$, we can effectively assume that density is uniform\footnote{For work going beyond this approximation in non-uniform condensates, see \cite{fetter2004kelvin}.}. Under this assumption, followed by considering a situation where the amplitude of the displacement is small, we have the standard result quoted in superfluid literature i.e. $\omega(k) \approx \frac{1}{2} k^2 \log \frac{1}{|k| a(R)}$, 
where $a(R) = \frac{1}{\sqrt{\mu}}\frac{1}{\sqrt{1-\frac{R^2}{R_{TF}^2}}}$ is the vortex core size via \eqref{cutoff-improved}. We remark that the boundary conditions on the string should quantize $k \sim \frac{n}{R_{TF}}$, so there is an approximate continuum of such operators above the gap to create a single vortex string. The spacing of these modes and exact dimensions are only visible at higher orders in the $Q$ expansion.

%

\section{Multi-Vortex Profile}\label{sec:multi-vortex}

Consider a collection of $N_v$ vortices at locations $\vec{R}_i$ with winding numbers $s_i$. The velocity field of such a contribution is additive and described by:
\be \label{multiple-vortex-velocity}
\vec{v}=\sum_i \vec{v}_i = \sum_i s_i \frac{\hat{z}\times (\vec{r}-\vec{R}_i)}{|\vec{r}-\vec{R}_i|^2} \implies \nabla \times \vec{v} = \sum_i s_i \delta(\vec{r}-\vec{R}_i)
\ee
Because the angular momentum is linear in the velocity field, this implies the total angular momentum of the system is given by the sum of the individual ones:
\begin{align} \label{angular-momentum-multi-vortex}
L &= \sum_i L_i = 
\begin{cases} 
 2\pi c_0 \left(\frac{\mu}{\omega}\right)^{2} \underset{i}{\sum}s_i \left(1- \frac{R_i^2}{R_{TF}^2}\right)^2~~~~~~ d=2  \\
\frac{5 \pi c_0}{8\sqrt{2}}   \left(\frac{\mu}{\omega}\right)^{3} \underset{i}{\sum} s_i \left(1-\frac{R_i^2}{R_{TF}^2}\right)^3 ~~~~~~d=3
\end{cases}
\end{align}
For vortices far from the boundary, where $\frac{R_{TF}-R_i}{R_{TF}} \sim \mathcal{O}(1)$ (as opposed to $Q$ suppressed number), we have that $L_i \sim s_i Q$.

We can compute the energy of a generic multi-vortex configuration explicitly from this velocity field \eqref{multiple-vortex-velocity}. The energy breaks up into single-vortex contributions and pair-wise interaction energies:
\be \label{energy-multi-vortex-schematic}
\Delta E = \frac{1}{2}\int \text{d}^dx ~n v^2 = \sum_i E_i + \sum_{i \neq j} \sum_j E_{ij} ~~~~
\ee
where the single vortex energy is already computed as
\be \label{self-energy-multi}
E_i = \frac{1}{2}\int \text{d}^dx ~n v_i^2 =
\begin{cases}
\omega\sqrt{\frac{c_0\pi}{2}} s_i^2 \sqrt{L_i} \log L_i  ~~~~~~d=2\\
\omega \left(\frac{5\pi^4 c_0}{8\sqrt{2}}\right)^{1/3} s_i^2  L_i^{2/3}\log L_i~~~~d=3
\end{cases}
\ee
and $E_{ij}$ is the interaction energy given by:
\be \label{interaction-energy-def}
E_{ij} = \int \text{d}^dx ~ n v_i \cdot v_j
\ee

In $d=2$ this integral evaluates to:
\begin{align} \label{interaction-energy-2d} 
\frac{E_{ij}}{s_i s_j} &= \pi c_0 \mu \left(1- \frac{\vec{R}_i \cdot \vec{R}_j}{R^2_{TF}}\right) \log\left(\frac{R^4_{TF}+R^2_i R^2_j - 2 R^2_{TF} \vec{R}_i \cdot \vec{R}_j}{(\vec{R}_i - \vec{R}_j)^4}\right)\\ \nonumber
 &- 2\pi c_0 \mu \frac{1}{R_{TF}^2} \left(R^2_i + R^2_j - R^2_{TF}\right)\\ \nonumber
 &+ 2\pi c_0 \mu \frac{1}{R_{TF}^2}|\vec{R}_i \times \vec{R}_j| \arctan
\left(\frac{|\vec{R}_i \times \vec{R}_j|}{R^2_{TF}-\vec{R}_i \cdot \vec{R}_j}\right) 
\end{align}
where $\vec{R}_i$ and $\vec{R}_j$ are the positions of the vortex pair with $R_j > R_i$ assumed without loss of generality. To leading order in the charge and small vortex separation this simplifies to:
\be \label{interaction-energy-2d-simple}
E_{ij} \sim s_i s_j \mu \log \frac{R_{TF}}{|\vec{R}_i - \vec{R}_j|}  + ...
\ee
This piece is the result of the singular nature of the vortices and describes their interaction. The analogous result of \eqref{interaction-energy-2d} for $d=3$ is not analytically tractable, but the leading interaction piece in the charge and small vortex separation is given by:
\be \label{interaction-energy-3d-simple}
E_{ij} \sim s_i s_j \mu^{\frac{3}{2}} R_{TF}\log \frac{R_{TF}}{|\vec{R}_i-\vec{R}_j|} + \cdots
\ee

One can extract several physical features of the multivortex profile using the expressions above for the energy. First of all, the minimum energy configurations per angular momentum will have $s_i = 1$ for every vortex as the energy scales quadratically in the charge but the angular momentum only scales linearly. The angular momentum for the entire configuration then scales as $L \sim N_v Q$ assuming $\frac{R_{TF}-R_i}{R_{TF}} \sim \mathcal{O}(1)$. Secondly, we remark that the logarithmic terms \eqref{interaction-energy-2d-simple} and \eqref{interaction-energy-3d-simple} imply that the minimal energy configuration will generically be a triangular array of vortices\cite{tkachenko1966vortex,campbell1979vortex}. Empirically this structure persists as the number of vortices is made large, even in the presence of a harmonic trap\cite{zwierlein2005vortices}.


In principle the energy, and therefore the operator dimension, should be found by fixing the angular momentum and varying over the positions $R_i$ to find the minimum energy configuration. However, for $N_v \sim \mathcal{O}(1)$ the interaction is negligible and the energy will scale as $E \sim N_v E_v$ where $E_v$ is the energy of a single vortex placed in the center of the trap. To consider $L$ parametrically larger than $Q$ we must consider $N_v \gg 1$. While we cannot exactly analyze \eqref{energy-multi-vortex-schematic} in this limit, we are justified in approximating the vortex density as a continuous quantity, corroborated by the fact that in this limit the interaction energy dominates and has terms which go as $N_v^2 \mu^{\frac{d}{2}}R_{TF}^{d-2}\sim L^2/I$, where $I$ is the moment of inertia, given later by Eq.~\eqref{answers-intertia}. 

\paragraph*{Continuum Approximation:}

We can take advantage of the fact the vortices are dense to coarse grain \eqref{multiple-vortex-velocity} and replace it with a continuous velocity field which satisfies:
\be \label{average-vorticity}
\oint_{C} \vec{v}\cdot \text{d}\vec{\ell} = 2\pi N_v(C)
\ee
where $N_v(C)$ is the number of vortices in the area enclosed by the curve $C$. Let $L$ be the angular momentum (to be precise the $z$ component of the angular momentum) of the configuration. We take a varational approach, minimizing the energy over smooth $v$ with fixed $L$. To this end, define:
\begin{align} \label{lagrange-E}
E_{\Omega} &= \frac{1}{2}\int \text{d}^dx ~ n v^2 - \Omega\left(\int \text{d}^dx ~ n \left(\vec{r} \times \vec{v}\right) \cdot \hat{z} - L\right) \\ &= \frac{1}{2}\int \text{d}^dx ~n \left(v - \Omega \hat{z}\times \vec{r}\right)^2 - \frac{\Omega^2}{2} \int \text{d}^dx ~n r^2 + \Omega L
\end{align}
where $\Omega$ is a Lagrange multiplier to fix the angular momentum. From \eqref{lagrange-E}, we can see that the minimum energy velocity field is that of a rotating rigid body with uniform vortex density:
\be \label{velocity-rigid-body}
\vec{v}= \Omega \hat{z} \times \vec{r} \implies \Delta E = \frac{\Omega^2}{2}\int \text{d}^dx~n r^2 = \frac{\Omega^2}{2}I
\ee
where $I$ is the moment of inertia of the condensate, computed from the density as:
\be \label{moment-of-inertia}
I=\int \text{d}^dx ~ n(r) r^2\,.
\ee
and $\Omega$ can be determined via its relation to $L$ as $\Omega = \frac{L}{I}$. Now the moment of inertia $I$ evaluates to 
\be \label{answers-intertia}
 I=\begin{cases}\frac{4}{3}\pi c_0 \frac{\mu^3}{\omega^4}= \frac{1}{\omega}\left(\frac{2}{3}\frac{1}{\sqrt{2\pi c_0}} Q^{\frac{3}{2}}\right) ~~ d=2\\ \\
\frac{5 \pi^2}{8 \sqrt{2}} c_0 \frac{\mu^4}{\omega^5} =  \frac{1}{\omega}\left(\frac{25}{1024}\left(\frac{25\pi^4}{32c_0^2}\right)^{1/6}Q^{\frac{4}{3}}\right)~~d=3
\end{cases}
\ee
Using \eqref{average-vorticity} and \eqref{velocity-rigid-body} we can also determine that the angular momentum of the configuration scales as $L\sim N_v Q$ as expected from \eqref{angular-momentum-multi-vortex}. Consequently, the energy is that of a rigid body with angular momentum $L$ and is given by:
\be \label{rigid-energy}
\Delta E = \frac{L^2}{2I}\,,
\ee
Notice that this leading order result is independent of the trap and the inhomogeneity of the density. Corrections will arise from the inhomogeneity of the trap and the discreteness of the vortices, but they are subleading in $N_v$ and suppressed in $R_{TF}$ \cite{sheehy2004vortices}. Indeed, that there are terms in the energy which scale as $N_v$ being neglected is visible in \eqref{energy-multi-vortex-schematic}.


We remark that there are constraints of the vortex density of the system. The vortex spacing $\lambda$ should be larger than the vortex core size i.e. $\lambda \gg a \sim \frac{1}{\sqrt{\mu_{eff}}}$. Beyond this limit we expect interactions to be strong and the EFT description to break down \cite{cooper2008rapidly}. Now, in a scenario where we have multiple vortices, a rough estimation yields that 
\begin{align}
N_v \sim \frac{R_{TF}^2}{\lambda^2}  \sim \begin{cases}
\sqrt{Q\ell}~~~&d=2\\
(Q\ell)^{1/3}~~~&d=3
\end{cases}
\end{align}
where $\ell$ is the typical angular momentum of a vortex in the multivortex configuration. Thus in $d=2$, the maximum angular momentum configularation that one can reach within the validity of the EFT corresponds to a maximum density of $N_v \sim Q$. Physically this means most of the vortices are near the center and $\ell \sim Q$ and the total angular momentum $L\sim Q^2$. For $d=3$ this corresponds to $N_v \sim Q^{2/3}$ which is less than $Q$ because the vortices are extended objects and the total angular momentum amounts to $L\sim Q^{5/3}$. But our EFT breaks down before this. Using particle vortex duality as in \ref{app:A}, one can see that the EFT breaks down when the electric field becomes comparable to magnetic field. This means that the EFT breaks down when the contribution coming from rigid body rotation becomes comparable to $\Delta_Q$. Hence, the maximum angular momentum that can be attained within the validity of our EFT is $L\sim Q^{3/2}$ in $d=2$ and $L\sim Q^{4/3}$ in $d=3$. 

These determine the absolute limits on the angular momentum accessible within our EFT and together with \eqref{rigid-energy} and \eqref{answers-intertia} imply the following operator dimension scaling:

\be \label{2d-op-rigid-body-scaling}
d=2 ~~~ \Delta_{Q,L} = \sqrt{\frac{9c_0\pi}{2}} \left( \frac{L^2}{Q^{3/2}}\right)+\Delta_Q ~~~~~ Q < L< Q^{3/2}
\ee

\be \label{3d-op-rigid-body-scaling}
d=3 ~~~ \Delta_{Q,L} = \frac{1024}{25}\left(\frac{32c_0^2}{25\pi^4}\right)^{1/6}\left(\frac{L^2}{Q^{4/3}}\right) +\Delta_Q ~~~~~ Q < L < Q^{4/3}
\ee

The above consitute the main results of this section.

\section{Conclusions and Future Directions}\label{sec:conclusion}

To summarize, we have calculated how the dimensions of operators in NRCFTs scale with number charge $Q$ and spin $L$ in the limit of $Q \gg 1$ via the state-operator correspondence. The NRCFTs under consideration exist in $d=2$ and $d=3$ and by assumption are described by the superfluid EFT. This allows for explicit calculations by studying phonon and vortex configurations of the superfluid. We expect applicability of our result to ``fermions at unitarity" and certain conformal anyon theories, as well any other NRCFT with this symmetry breaking behavior in its large charge sector\cite{nishida2007nonrelativistic,doroud2016superconformal,doroud2018conformal,Jackiw:1991au}. In fact the superfluid state of unitary fermions in a harmonic trap has been experimentally observed, including the formation of vortices \cite{zwierlein2005vortices}. 

The most direct extension of these results would be to go to beyond the leading order scaling. To do so would require reasoning about the divergences associated with the vortex core, the size and structure of which is entirely determined by UV physics. It should be possible to regulate such divergences by considering operators localized on the vortex. Such a procedure in the relativistic effective string theory was worked out in \cite{hellerman2017boundary,hellerman2015string} and the effective string theory of vortex lines in superfluids was explored in \cite{horn2015effective}. A similar analysis has also been applied to divergences of the superfluid EFT near $R_{TF}$, associated with the dilute regime of size $\delta$ \cite{Priv_Comm_Simeon} .

It would be especially interesting to study other possible symmetry breaking patterns, such as those relevant for chiral superfluids \cite{hoyos2014effective}. As mentioned in this large angular momentum regime the vortices are arranged as a triangular lattice. Deformations of this vortex lattice are a novel excitation in this limit, known as `Tkachenko modes' \cite{tkachenko1969elasticity}. Presumably these excited states would correspond to a tower of low-lying operators above the operator which creates the vortex-lattice. However, treatment these modes and corrections to the results \eqref{2d-op-rigid-body-scaling} and \eqref{3d-op-rigid-body-scaling} would require us to think about a new EFT which captures the spontaneous breaking of spatial symmetry by the vortex lattice. This EFT has been worked out by \cite{moroz2018effective} and may be adaptable to the Schr\"odinger invariant case in a trap. Especially interesting would be systems with a Fermi surface, however such a critical state must necessarily be a non-fermi liquid following the results of \cite{rothstein2018symmetry}. 

Another interesting direction would be to consider correlation functions of charged spinning operators in these NRCFTs. The universal scaling of the 3-point function and higher are all explicitly calculable within this EFT, as was done for scalar charged operators in \cite{Kravec:2018qnu}. In relativistic CFTs this was worked out in \cite{Cuomo:2017vzg,monin2017semiclassics} for certain operators. We leave this and other questions for future work. 

\section*{Acknowledgements}
The authors acknowledge useful comments from John McGreevy. This work was in part supported by the US Department of Energy (DOE) under cooperative research agreement DE-SC0009919. SP acknowledges partial support from the Inamori Fellowship, funded by Inamori Foundation. 

\appendix

\section{Particle-Vortex Duality}\label{app:A}
Here we briefly review the particle vortex duality in nonrelativistic set up. The aim of the appendix is to cast the vortex dynamics in terms of an electrostatic (in $d=3$ the gauge field is $2$ form field, hence we coin the term ``gaugostatic") problem, leveraging the duality. The idea is to solve the gaugostatic problem to figure out the field strength, which in turn gives us the velocity profile of the vortex, again using the dictionary of duality.

We consider the leading order superfluid Lagrangian in the presence of a potential $A_0=\frac{1}{2}\omega^2 r^2$, $A_i=0$:
\be \label{leading-order}
\mathcal{L}= c_0 X^{\frac{d+2}{2}} \equiv P(X) ~~~~~ X\equiv \partial_0 \chi -  A_0 - \frac{1}{2}(\partial_i \chi)^2
\ee
The number density and superfluid velocity are defined respectively as:
\be \label{number-velocity}
n= \frac{\partial L}{\partial \dot{\chi}}= c_0 \left(\frac{d}{2}+1\right) X^{\frac{d}{2}} ~~~~~ v_i = - \partial_i \chi
\ee
The action \eqref{leading-order} has a $U(1)$ symmetry of $\chi \rightarrow \chi + c$ whose current can be written as:
\be \label{u1-current}
j^\mu = \( n , n v^i \)
\ee
For simplicity and physical relevance, we'll focus on the cases of $d=2$ and $d=3$. In $d=2$, we can define:
\be \label{2d-duality}
j^\mu = \epsilon^{\mu \nu \rho} \partial_\nu a_\rho = \frac{1}{2} \epsilon^{\mu \nu \rho} f_{\nu \rho}
\ee
for a one-form gauge field $a_\mu$ and field strength $f_{\mu \nu} = \partial_\mu a_\nu - \partial_\nu a_\mu$. This relates the superfluid variables \eqref{number-velocity} to the dual electric and magnetic fields as:
\be \label{2d-number-velocity-dual}
n = \epsilon^{ij} \partial_i a_j \equiv b ~~~~~ v^i = \frac{ \epsilon^{ij} f_{0j}}{b} \equiv \frac{ \epsilon^{ij} e_j}{b}
\ee
Similarly, in $d=3$ we'll define the current in terms of a dual two-form gauge field $B_{\mu \nu}$ 
\be \label{3d-duality}
j^\mu = \epsilon^{\mu \nu \rho \sigma} \partial_\nu B_{\rho \sigma} = \frac{1}{3} \epsilon^{\mu \nu \rho \sigma} G_{\nu \rho \sigma}
\ee
where $G_{\mu \nu \rho}=\partial_\mu B_{\nu \rho} + \partial_\nu B_{\rho \mu} + \partial_{\rho} B_{\mu \nu}$ is the three-form field strength. The superfluid variables are then expressible as:
\be \label{3d-number-velocity-dual}
n = \epsilon^{ijk} \partial_i B_{jk} \equiv Y ~~~~~ v^i = \frac{1}{3}\frac{ \epsilon^{ijk} G_{0jk}}{Y} 
\ee
Vortices act as sources for the gauge fields and couple minimally as:
\be \label{vortex-coupling}
d=2: ~~~ J^V_\mu a^\mu ~~~~~~~~~ d=3: ~~~ \frac{1}{4}J^V_{\mu \nu} B^{\mu \nu}
\ee
To implement the duality transformation, we note that internal energy $\epsilon(n)$ is given by $ nX - P(X)$ and so we can rewrite the \eqref{leading-order} as 
\begin{align}
\mathcal{L}&=nX-\epsilon(n)= n\left(\dot{\chi}-A_0-\frac{1}{2}(\partial_i\chi)(\partial^i\chi)\right)-\epsilon(n)\\
&=\frac{1}{2}nv^2-\epsilon(n)+n\left(\dot{\chi}+v^i\partial_i\chi\right)
\end{align}
where we have used $v_i=-\partial_i\chi$ and $n$ is understood as a function of $\chi$ and its derivatives.

The internal energy is given by:
\be \label{internal-energy}
d=2: ~~~ \epsilon(n)= \frac{1}{4c_0}n^2 ~~~~~~~~~~ d=3: ~~~ \epsilon(n) = \frac{3}{5} \left(\frac{2}{5c_0}\right)^{\frac{2}{3}}  n^{\frac{5}{3}}
\ee
Using the relation \eqref{2d-number-velocity-dual} we can express the Lagrangian in $d=2$ as:
\be \label{dual-Lagrangian-2d}
\mathcal{L}= \frac{1}{2}\frac{e^2}{b}- \frac{1}{4c_0}b^2 - b A_0
\ee
This equation describes a kind of non-linear electrodynamics with a modified Gauss law:
\be \label{2d-gauss}
\partial_i \left(\frac{e^i}{b}\right) = J^V_0
\ee
Similarly the Lagrangian in $d=3$ is given via \eqref{3d-number-velocity-dual} as:
\be \label{dual-Lagrangian-3d}
\mathcal{L}= \frac{1}{9} \frac{G_{0ij}G_0{}^{ij}}{Y} -\frac{3}{5} \left(\frac{2}{5c_0}\right)^{\frac{2}{3}}  Y^{\frac{5}{3}}- Y A_0
\ee
with a ``Gauss law" of:
\be \label{3d-gauss}
\partial^i \left(\frac{G_{0ij}}{Y}\right) = J^V_{j0}
\ee

Now consider a motion of charged particle under the gauge field, sourced by $J^V$ . In what follows, we will show that to leading order we can treat this as a ``gaugostatic" problem and the velocity $V_i$ of the charged particle is negligible. If $V_i$ is negligible, one can potentially drop the kinetic term in the Lagrangian. As a result, the equation of motion for the particle turns out to be the one where there is no Lorentz force acting on the particle. This implies that $V_i$ is of the same order as $|e|/b$ (in $d=3$, this is $ \frac{\sqrt{G_{0ij}G_{0}{}^{ij}}}{Y} $). For self consistency, we need to ensure $V_i$ is very small i.e. the ratio $|e|/b$ is very small. This helps us to render the problem of vortex dynamics into a problem of ``gaugostatics". In order to do that, we linearize \eqref{dual-Lagrangian-2d} and $\eqref{dual-Lagrangian-3d}$ around parametrically large magnetic field $b$ and $Y$ and we see that the coupling goes as
$b$ in $d=2$ and in $d=3$, this goes like $Y$. Hence the electric field strength $|e|$ in $d=2$ and $\sqrt{G_{0ij}G_{0}{}^{ij}}$ in $d=3$ goes like $\sqrt{b}$ and $\sqrt{Y}$ respectively and we have 
\begin{align}
\sqrt{V_iV^i}\sim \frac{|e|}{b} \sim \frac{1}{\sqrt{Q}}\,, \text{in}\, d=2\\
\sqrt{V_iV^i}\sim \frac{\sqrt{G_{0ij}G_{0}{}^{ij}}}{Y} \sim \frac{1}{\sqrt{Q}}\,, \text{in}\, d=3
\end{align}
Thus it is self consistent to assume that the charged particle is just drifting without any Lorentz force acting on it. 
\section{A Contour integral}\label{app:B}
This appendix contains the evauation of contour integrals, needed to figure out the vortex interaction energy in the multivortex scenario. In $d=2$, the vortex interaction energy goes like
\begin{align}
\int \text{d}r\ rn(r)\int \text{d}\theta\ \vec{v}_{i}\cdot \vec{v}_j 
\end{align}
wherre as for $d=3$, we have an extra integral along the $z$ axis and $r$ becomes the radius in cylindrical coordinate.  In both cases, the $\theta$ integral can be done using contour integral and expressing $\vec{v}_{i}$ in terms of  complex variables given by 
\begin{align}
v_i=\frac{i}{\bar z- \bar z_{i}}\,,&\ v_i^{*}=\frac{-i}{z-  z_{i}}
\end{align}

Hence the integral evaluates to 
\begin{align}
I=\int\text{d}\theta\ \vec{v}_i\cdot \vec{v}_j &= Re\left( \int\text{d}z\ \frac{-i}{z} v_{i}v_{j}^{*}\right)\\
&= Re\left( \int\text{d}z\ \frac{-i}{z} \frac{1}{(\bar z- \bar z_{i})(z-z_j)}\right)
\end{align}

Now we note that $z\bar z=r^2$ and $z_i\bar z_i=R_i^2$ to rewrite the integral in following manner: 
\begin{align}
I&= Re\left( \int\text{d}z\ -i \frac{z_i}{(r^2 z_i- R_i^2z)(z-z_j)}\right)=Re\left( \int\text{d}z\ \frac{-i}{-R_i^2} \frac{z_i}{\left(z-\frac{r^2}{R_i^2} z_i\right)(z-z_j)}\right)
\end{align}

The poles are located at $z=z_j\,,z=\frac{r^2}{R_i^2} z_i$ i.e. they lie on the circle of radius $|z_j|=R_j$ and $|\frac{r^2}{R_i^2} z_i|=\frac{r^2}{R_i}$. \\

Without loss of generality, we  consider $R_i<R_j$. Now there can be three scenarios: 
\begin{enumerate}
\item $r<R_i<R_j$ implies $ \frac{r^{2}}{R_i}< r<R_j$, hence the pole at $z=\frac{r^2}{R_i^2} z_i$ is picked, answer is
\begin{align}
I=Re \left( -2\pi \frac{1}{r^2-z_jz_i^*} \right)= -\frac{2\pi}{r^4+R_j^2R_i^2-2r^2R_iR_j\cos(\phi)}\left(r^2-R_iR_j\cos(\phi)\right)
\end{align}
\item $R_i<R_j<r$ implies $R_j<r<\frac{r^2}{R_i}$, hence the pole at $z=z_j$ is picked. and the answer is 
\begin{align}
I=\frac{2\pi}{r^4+R_j^2R_i^2-2r^2R_iR_j\cos(\phi)}\left(r^2-R_iR_j\cos(\phi)\right)
\end{align}
\item $R_i<r<R_j$ implies $r<R_j$ and $r<\frac{r^2}{R_i}$, so none of the poles is picked, the answer is $0$. 
\end{enumerate}

Summing up we can write
\begin{align}
I= \frac{\pi\left(r^2-R_iR_j\cos(\phi)\right)}{r^4+R_j^2R_i^2-2r^2R_iR_j\cos(\phi)} \left[sgn(r-R_i)+sgn(r-R_j)\right]
\end{align}
{\bibliographystyle{bibstyle2017}

\bibliography{references}

\providecommand{\href}[2]{#2}\begingroup\begin{thebibliography}{10}

\bibitem{donnelly1991quantized}
R.~J. Donnelly, {\it Quantized vortices in helium ii}, , vol.~2.
\newblock Cambridge University Press, 1991.

\bibitem{vitiello1996vortex}
S.~Vitiello, L.~Reatto, G.~Chester, and M.~Kalos, {\it Vortex line in
  superfluid he 4: A variational monte carlo calculation},  {\sf Physical
  Review B} {\sf {54} }{\sf no.~2, }{\sf (1996) }{\sf 1205}.

\bibitem{ortiz1995core}
G.~Ortiz and D.~M. Ceperley, {\it Core structure of a vortex in superfluid he
  4},  {\sf Physical review letters} {\sf {75} }{\sf no.~25, }{\sf (1995) }{\sf
  4642}.

\bibitem{giorgini1996vortex}
S.~Giorgini, J.~Boronat, and J.~Casulleras, {\it Vortex excitation in
  superfluid 4 he: A diffusion monte carlo study},  {\sf Physical review
  letters} {\sf {77} }{\sf no.~13, }{\sf (1996) }{\sf 2754}.

\bibitem{baym1969superfluidity}
G.~Baym, C.~Pethick, and D.~Pines, {\it Superfluidity in neutron stars},  {\sf
  Nature} {\sf {224} }{\sf no.~5220, }{\sf (1969) }{\sf 673}.

\bibitem{hartnoll2016holographic}
S.~A. Hartnoll, A.~Lucas, and S.~Sachdev, {\it Holographic quantum matter},
  {\sf arXiv preprint arXiv:1612.07324} {\sf (2016) }.

\bibitem{hellerman2015cft}
S.~Hellerman, D.~Orlando, S.~Reffert, and M.~Watanabe, {\it On the cft operator
  spectrum at large global charge},  {\sf Journal of High Energy Physics} {\sf
  {2015} }{\sf no.~12, }{\sf (2015) }{\sf 1--34}.

\bibitem{hellerman2017note}
S.~Hellerman, N.~Kobayashi, S.~Maeda, and M.~Watanabe, {\it A note on
  inhomogeneous ground states at large global charge},  {\sf arXiv preprint
  arXiv:1705.05825} {\sf (2017) }.

\bibitem{monin2017semiclassics}
A.~Monin, D.~Pirtskhalava, R.~Rattazzi, and F.~K. Seibold, {\it Semiclassics,
  goldstone bosons and cft data},  {\sf Journal of High Energy Physics} {\sf
  {2017} }{\sf no.~6, }{\sf (2017) }{\sf 11}.

\bibitem{banerjee2018conformal}
D.~Banerjee, S.~Chandrasekharan, and D.~Orlando, {\it Conformal dimensions via
  large charge expansion},  {\sf Physical review letters} {\sf {120} }{\sf
  no.~6, }{\sf (2018) }{\sf 061603}.

\bibitem{de2018large}
A.~de~la Fuente, {\it The large charge expansion at large n},  {\sf arXiv
  preprint arXiv:1805.00501} {\sf (2018) }.

\bibitem{Jafferis:2017zna}
D.~Jafferis, B.~Mukhametzhanov, and A.~Zhiboedov, {\it {Conformal Bootstrap At
  Large Charge}},  \href{http://dx.doi.org/10.1007/JHEP05(2018)043}{{\sf JHEP}
  {\sf {05} }{\sf (2018) }{\sf 043}},
\href{http://arxiv.org/abs/1710.11161}{{\ttfamily arXiv:1710.11161 [hep-th]}}.

\bibitem{Cuomo:2017vzg}
G.~Cuomo, A.~de~la Fuente, A.~Monin, D.~Pirtskhalava, and R.~Rattazzi, {\it
  {Rotating superfluids and spinning charged operators in conformal field
  theory}},  \href{http://dx.doi.org/10.1103/PhysRevD.97.045012}{{\sf Phys.
  Rev.} {\sf {D97} }{\sf no.~4, }{\sf (2018) }{\sf 045012}},
\href{http://arxiv.org/abs/1711.02108}{{\ttfamily arXiv:1711.02108 [hep-th]}}.

\bibitem{zwierlein2005vortices}
M.~W. Zwierlein, J.~R. Abo-Shaeer, A.~Schirotzek, C.~H. Schunck, and
  W.~Ketterle, {\it Vortices and superfluidity in a strongly interacting fermi
  gas},  {\sf Nature} {\sf {435} }{\sf no.~7045, }{\sf (2005) }{\sf 1047}.

\bibitem{kobach:2018nmt}
A.~Kobach and S.~Pal, {\it {Conformal Structure of the Heavy Particle EFT
  Operator Basis}},
  \href{http://dx.doi.org/10.1016/j.physletb.2018.06.060}{{\sf Phys. Lett.}
  {\sf {B783} }{\sf (2018) }{\sf 311--319}},
\href{http://arxiv.org/abs/1804.01534}{{\ttfamily arXiv:1804.01534 [hep-ph]}}.

\bibitem{Henkel:2003pu}
M.~Henkel and J.~Unterberger, {\it {Schrodinger invariance and space-time
  symmetries}},  \href{http://dx.doi.org/10.1016/S0550-3213(03)00252-9}{{\sf
  Nucl. Phys.} {\sf {B660} }{\sf (2003) }{\sf 407--435}},
\href{http://arxiv.org/abs/hep-th/0302187}{{\ttfamily arXiv:hep-th/0302187
  [hep-th]}}.

\bibitem{Regal:2004zza}
C.~A. Regal, M.~Greiner, and D.~S. Jin, {\it {Observation of Resonance
  Condensation of Fermionic Atom Pairs}},
\href{http://dx.doi.org/10.1103/PhysRevLett.92.040403}{{\sf Phys. Rev. Lett.}
  {\sf {92} }{\sf (2004) }{\sf 040403}}.

\bibitem{Zwierlein:2004zz}
M.~W. Zwierlein, C.~A. Stan, C.~H. Schunck, S.~M.~F. Raupach, A.~J. Kerman, and
  W.~Ketterle, {\it {Condensation of Pairs of Fermionic Atoms near a Feshbach
  Resonance}},
\href{http://dx.doi.org/10.1103/PhysRevLett.92.120403}{{\sf Phys. Rev. Lett.}
  {\sf {92} }{\sf (2004) }{\sf 120403}}.

\bibitem{Kaplan:1998tg}
D.~B. Kaplan, M.~J. Savage, and M.~B. Wise, {\it {A New expansion for
  nucleon-nucleon interactions}},
  \href{http://dx.doi.org/10.1016/S0370-2693(98)00210-X}{{\sf Phys. Lett.} {\sf
  {B424} }{\sf (1998) }{\sf 390--396}},
\href{http://arxiv.org/abs/nucl-th/9801034}{{\ttfamily arXiv:nucl-th/9801034
  [nucl-th]}}.

\bibitem{Kaplan:1998we}
D.~B. Kaplan, M.~J. Savage, and M.~B. Wise, {\it {Two nucleon systems from
  effective field theory}},
  \href{http://dx.doi.org/10.1016/S0550-3213(98)00440-4}{{\sf Nucl. Phys.} {\sf
  {B534} }{\sf (1998) }{\sf 329--355}},
\href{http://arxiv.org/abs/nucl-th/9802075}{{\ttfamily arXiv:nucl-th/9802075
  [nucl-th]}}.

\bibitem{Chin:2001uan}
C.~Chin, V.~Vuletić, A.~J. Kerman, and S.~Chu, {\it {High precision Feshbach
  spectroscopy of ultracold cesium collisions}},
\href{http://dx.doi.org/10.1016/S0375-9474(01)00461-4}{{\sf Nucl. Phys.} {\sf
  {A684} }{\sf (2001) }{\sf 641--645}}.

\bibitem{Roberts:1998zz}
J.~L. Roberts, N.~R. Claussen, J.~P. Burke, C.~H. Greene, E.~A. Cornell, and
  C.~E. Wieman, {\it {Resonant Magnetic Field Control of Elastic Scattering in
  Cold R-85b}},
\href{http://dx.doi.org/10.1103/PhysRevLett.81.5109}{{\sf Phys. Rev. Lett.}
  {\sf {81} }{\sf (1998) }{\sf 5109--5112}}.

\bibitem{loftus2002resonant}
T.~Loftus, C.~Regal, C.~Ticknor, J.~Bohn, and D.~S. Jin, {\it Resonant control
  of elastic collisions in an optically trapped fermi gas of atoms},  {\sf
  Physical review letters} {\sf {88} }{\sf no.~17, }{\sf (2002) }{\sf 173201}.

\bibitem{Chen:2017tij}
X.~Chen, E.~Fradkin, and W.~Witczak-Krempa, {\it {Gapless quantum spin chains:
  multiple dynamics and conformal wavefunctions}},
  \href{http://dx.doi.org/10.1088/1751-8121/aa8dbc}{{\sf J. Phys.} {\sf {A50}
  }{\sf no.~46, }{\sf (2017) }{\sf 464002}},
\href{http://arxiv.org/abs/1707.02317}{{\ttfamily arXiv:1707.02317
  [cond-mat.str-el]}}.

\bibitem{Mehen:1999nd}
T.~Mehen, I.~W. Stewart, and M.~B. Wise, {\it {Conformal invariance for
  nonrelativistic field theory}},
  \href{http://dx.doi.org/10.1016/S0370-2693(00)00006-X}{{\sf Phys. Lett.} {\sf
  {B474} }{\sf (2000) }{\sf 145--152}},
\href{http://arxiv.org/abs/hep-th/9910025}{{\ttfamily arXiv:hep-th/9910025
  [hep-th]}}.

\bibitem{Nishida:2010tm}
Y.~Nishida and D.~T. Son, {\it {Unitary Fermi gas, epsilon expansion, and
  nonrelativistic conformal field theories}},
  \href{http://dx.doi.org/10.1007/978-3-642-21978-8_7}{{\sf Lect. Notes Phys.}
  {\sf {836} }{\sf (2012) }{\sf 233--275}},
\href{http://arxiv.org/abs/1004.3597}{{\ttfamily arXiv:1004.3597
  [cond-mat.quant-gas]}}.

\bibitem{Nishida:2007pj}
Y.~Nishida and D.~T. Son, {\it {Nonrelativistic conformal field theories}},
  \href{http://dx.doi.org/10.1103/PhysRevD.76.086004}{{\sf Phys. Rev.} {\sf
  {D76} }{\sf (2007) }{\sf 086004}},
\href{http://arxiv.org/abs/0706.3746}{{\ttfamily arXiv:0706.3746 [hep-th]}}.

\bibitem{goldberger2015ope}
W.~D. Goldberger, Z.~U. Khandker, and S.~Prabhu, {\it Ope convergence in
  non-relativistic conformal field theories},  {\sf Journal of High Energy
  Physics} {\sf {2015} }{\sf no.~12, }{\sf (2015) }{\sf 1--31}.

\bibitem{Golkar:2014mwa}
S.~Golkar and D.~T. Son, {\it {Operator Product Expansion and Conservation Laws
  in Non-Relativistic Conformal Field Theories}},
  \href{http://dx.doi.org/10.1007/JHEP12(2014)063}{{\sf JHEP} {\sf {12} }{\sf
  (2014) }{\sf 063}},
\href{http://arxiv.org/abs/1408.3629}{{\ttfamily arXiv:1408.3629 [hep-th]}}.

\bibitem{Pal:2018idc}
S.~Pal, {\it {Unitarity and universality in nonrelativistic conformal field
  theory}},  \href{http://dx.doi.org/10.1103/PhysRevD.97.105031}{{\sf Phys.
  Rev.} {\sf {D97} }{\sf no.~10, }{\sf (2018) }{\sf 105031}},
\href{http://arxiv.org/abs/1802.02262}{{\ttfamily arXiv:1802.02262 [hep-th]}}.

\bibitem{nishida2007nonrelativistic}
Y.~Nishida and D.~T. Son, {\it Nonrelativistic conformal field theories},  {\sf
  Physical Review D} {\sf {76} }{\sf no.~8, }{\sf (2007) }{\sf 086004}.

\bibitem{Kravec:2018qnu}
S.~M. Kravec and S.~Pal, {\it {Nonrelativistic Conformal Field Theories in the
  Large Charge Sector}},  \href{http://dx.doi.org/10.1007/JHEP02(2019)008}{{\sf
  JHEP} {\sf {02} }{\sf (2019) }{\sf 008}},
\href{http://arxiv.org/abs/1809.08188}{{\ttfamily arXiv:1809.08188 [hep-th]}}.

\bibitem{Favrod:2018xov}
S.~Favrod, D.~Orlando, and S.~Reffert, {\it {The large-charge expansion for
  Schr\"odinger systems}},
\href{http://arxiv.org/abs/1809.06371}{{\ttfamily arXiv:1809.06371 [hep-th]}}.

\bibitem{son2006general}
D.~Son and M.~Wingate, {\it General coordinate invariance and conformal
  invariance in nonrelativistic physics: Unitary fermi gas},  {\sf Annals of
  Physics} {\sf {321} }{\sf no.~1, }{\sf (2006) }{\sf 197--224}.

\bibitem{pethick2008bose}
C.~J. Pethick and H.~Smith, {\it Bose--einstein condensation in dilute gases},
  .
\newblock Cambridge university press, 2008.

\bibitem{bruun2001vortex}
G.~Bruun and L.~Viverit, {\it Vortex state in superfluid trapped fermi gases at
  zero temperature},  {\sf Physical Review A} {\sf {64} }{\sf no.~6, }{\sf
  (2001) }{\sf 063606}.

\bibitem{sheehy2004vortices}
D.~E. Sheehy and L.~Radzihovsky, {\it Vortices in spatially inhomogeneous
  superfluids},  {\sf Physical Review A} {\sf {70} }{\sf no.~6, }{\sf (2004)
  }{\sf 063620}.

\bibitem{groszek2018motion}
A.~J. Groszek, D.~M. Paganin, K.~Helmerson, and T.~P. Simula, {\it Motion of
  vortices in inhomogeneous bose-einstein condensates},  {\sf Physical Review
  A} {\sf {97} }{\sf no.~2, }{\sf (2018) }{\sf 023617}.

\bibitem{fetter2004kelvin}
A.~L. Fetter, {\it Kelvin mode of a vortex in a nonuniform bose-einstein
  condensate},  {\sf Physical Review A} {\sf {69} }{\sf no.~4, }{\sf (2004)
  }{\sf 043617}.

\bibitem{tkachenko1966vortex}
V.~Tkachenko, {\it On vortex lattices},  {\sf Sov. Phys. JETP} {\sf {22} }{\sf
  no.~6, }{\sf (1966) }{\sf 1282--1286}.

\bibitem{campbell1979vortex}
L.~Campbell and R.~M. Ziff, {\it Vortex patterns and energies in a rotating
  superfluid},  {\sf Physical Review B} {\sf {20} }{\sf no.~5, }{\sf (1979)
  }{\sf 1886}.

\bibitem{cooper2008rapidly}
N.~R. Cooper, {\it Rapidly rotating atomic gases},  {\sf Advances in Physics}
  {\sf {57} }{\sf no.~6, }{\sf (2008) }{\sf 539--616}.

\bibitem{doroud2016superconformal}
N.~Doroud, D.~Tong, and C.~Turner, {\it On superconformal anyons},  {\sf
  Journal of High Energy Physics} {\sf {2016} }{\sf no.~1, }{\sf (2016) }{\sf
  138}.

\bibitem{doroud2018conformal}
N.~Doroud, D.~Tong, and C.~Turner, {\it The conformal spectrum of non-abelian
  anyons},  {\sf SciPost Physics} {\sf {4} }{\sf no.~4, }{\sf (2018) }{\sf
  022}.

\bibitem{Jackiw:1991au}
R.~Jackiw and S.-Y. Pi, {\it {Selfdual Chern-Simons solitons}},
  \href{http://dx.doi.org/10.1143/PTPS.107.1}{{\sf Prog. Theor. Phys. Suppl.}
  {\sf {107} }{\sf (1992) }{\sf 1--40}}.
[,465(1991)].

\bibitem{hellerman2017boundary}
S.~Hellerman and I.~Swanson, {\it Boundary operators in effective string
  theory},  {\sf Journal of High Energy Physics} {\sf {2017} }{\sf no.~4, }{\sf
  (2017) }{\sf 85}.

\bibitem{hellerman2015string}
S.~Hellerman and I.~Swanson, {\it String theory of the regge intercept},  {\sf
  Physical review letters} {\sf {114} }{\sf no.~11, }{\sf (2015) }{\sf 111601}.

\bibitem{horn2015effective}
B.~Horn, A.~Nicolis, and R.~Penco, {\it Effective string theory for vortex
  lines in fluids and superfluids},  {\sf Journal of High Energy Physics} {\sf
  {2015} }{\sf no.~10, }{\sf (2015) }{\sf 153}.

\bibitem{Priv_Comm_Simeon}
S.~Hellerman. {Private Communication}, 2019.

\bibitem{hoyos2014effective}
C.~Hoyos, S.~Moroz, and D.~T. Son, {\it {Effective theory of chiral
  two-dimensional superfluids}},
  \href{http://dx.doi.org/10.1103/PhysRevB.89.174507}{{\sf Phys. Rev.} {\sf
  {B89} }{\sf no.~17, }{\sf (2014) }{\sf 174507}},
\href{http://arxiv.org/abs/1305.3925}{{\ttfamily arXiv:1305.3925
  [cond-mat.quant-gas]}}.

\bibitem{tkachenko1969elasticity}
V.~Tkachenko, {\it Elasticity of vortex lattices},  {\sf Soviet Journal of
  Experimental and Theoretical Physics} {\sf {29} }{\sf (1969) }{\sf 945}.

\bibitem{moroz2018effective}
S.~Moroz, C.~Hoyos, C.~Benzoni, and D.~T. Son, {\it {Effective field theory of
  a vortex lattice in a bosonic superfluid}},
\href{http://arxiv.org/abs/1803.10934}{{\ttfamily arXiv:1803.10934
  [cond-mat.quant-gas]}}.

\bibitem{rothstein2018symmetry}
I.~Z. Rothstein and P.~Shrivastava, {\it Symmetry realization via a dynamical
  inverse higgs mechanism},  {\sf Journal of High Energy Physics} {\sf {2018}
  }{\sf no.~5, }{\sf (2018) }{\sf 14}.

\end{thebibliography}\endgroup

\hypersetup{urlcolor=RoyalBlue!60!black}
}

\end{document}